\documentclass{elsart}
\usepackage{amsmath}
\usepackage{amsfonts}
\usepackage{amssymb}
\usepackage{graphicx}

\begin{document}

\begin{frontmatter}

\title{Lagrangian Formalism in Perturbed Nonlinear  
   Klein-Gordon Equations}

%\title{Collective Coordinate Approach for Perturbed Nonlinear  
%   Klein-Gordon Equations}
  
\author[a,b]{Niurka R.\ Quintero\thanksref{cor}}
\thanks[cor]{Corresponding author. E-mail: niurka@euler.us.es}
and \author[a,c]
{El\'{\i}as Zamora-Sillero\thanksref{auth}}  
\thanks[auth]{E-mail: elias@euler.us.es}

\address[a]{Departamento de F\'{\i}sica Aplicada I,
Escuela Universitaria Polit\'ecnica,  
Universidad de Sevilla, Virgen de \'Africa 7, 41011, Sevilla,
Spain}
\address[b]{Instituto {\sc Carlos I} de F\'{\i}sica Te\'orica y
Computacional
Universidad de Granada. E-18071 Granada, Spain}
\address[c]{Departamento de An\'alisis Matem\'atico,
Facultad de Matem\'aticas, Universidad de Sevilla, 
Apartado de Correos 1160, Sevilla E-41080, Spain}

\journal{Physica D}

\date{\today}
\maketitle

\begin{abstract}
{We develop an alternative approach to 
study the effect of the generic 
perturbation (in addition to explicitly considering the loss term)
in the nonlinear Klein-Gordon equations. 
By a change of the 
variables that cancel the dissipation term 
we are able to write the Lagrangian density and then, 
calculate the Lagrangian as a function of collective 
variables. We use  
the Lagrangian formalism together with the Rice {\it Ansatz} to derive the equations  
of motion of the collective coordinates (CCs) for the 
perturbed sine-Gordon (sG)  
and $\phi^{4}$ systems. 
For the $N$ collective coordinates, regardless of the {\it Ansatz} used, 
we show that, for the nonlinear Klein-Gordon equations, 
this approach is equivalent to the {\it 
Generalized Traveling Wave Ansatz} ({\it GTWA}).       
%We show that 
%this method leads to the approximated solution more directly 
%that the previous known approaches. 
%We obtain the equations that 
%satisfy the collective coordinates for some particular choice 
%of the solution of the perturbed equations.
}   
\end{abstract}

\begin{keyword}
Collective coordinates \sep Solitons \sep Solitary waves \sep 
Perturbed Nonlinear Klein-Gordon equations.  

\PACS{03.40.Kf \sep 04.25.-g \sep 04.20.Fy}  
%03.40.Kf       Waves and wave propagation: general mathematical aspects 
%42.81.Dp       Propagation, scattering, and losses; solitons
%04.20.Fy       Canonical formalism, Lagrangians, and variational principles
%04.25.-g       Approximation methods; equations of motion
\end{keyword}

\end{frontmatter}

\date{\today}% It is always \today, today,
             %  but any date may be explicitly specified

\maketitle

\section{Introduction}\label{intro}

 The solitons and solitary waves under the external perturbations 
have been extensively studied in the last three decades 
(see e.g. \cite{KM,SIAM,Scott} and references there in). 
This has been more than justified 
%The reason of that is well justified 
since in the physical real systems, modeled by the equations 
related with these solutions, the dissipation and the external 
forces are unavoidably present \cite{fogel76,mc}. In this work 
we use the Lagrangian formalism to study the effect of damping 
and external forces  
in the nonlinear Klein-Gordon systems, and in particular we use as examples   
sine-Gordon (sG) and $\phi^{4}$ (this method can be extended to other  
nonlinear Klein-Gordon equations such as, the double sine-Gordon 
(DsG) \cite{Scott} or the asymmetric double sine-Gordon (ADSG) 
\cite{sn}). To illustrate the advantage of using this method 
over previous ones used in this kind of problem, let us 
revise the situation of the perturbation theory and collective 
coordinates (CCs in short) on the aforesaid systems. 
First of all, the authors of  \cite{fogel76} 
developed a technique to investigate the influence of external 
perturbations, such as damping, dc force and 
spatial inhomogeneity, on solitary waves. 
This method is based on the 
expansion of the solution of the 
perturbed problem in the complete set of eigenfunctions of the 
Sturm-Liouville problem associated with the linearized partial 
differential equation (PDE) around 
a kink solution. Later, the dynamic of the breather and 
kink-antikink solution in the perturbed sG equation was also 
considered using a new 
perturbative analysis \cite{mc,sal-scott,ariyasu}, 
where the radiation field was also included 
\cite{mc,sal-scott}. 
All these methods 
involve cumbersome calculations, even more so if we extend them to the 
$\phi^{4}$ equation where the internal mode is present. 
In particular, if it is assumed 
that the perturbations only alter the center of the kink, $X(t)$, 
the calculations can be simplified.  
In this case, it was shown that the perturbative method is 
equivalent to the variation of the energy of the system \cite{mc}. 
However, the presence of the internal mode in $\phi^{4}$ 
and 
some phonon's modes in sG are also able to change   
the kink's width, $l(t)$. Then, the solitary waves 
under certain perturbations can exhibit resonance's phenomena 
\cite{prl,rc,np} and it is not appropriate to only consider one 
collective coordinate, for example, the center of the kink, in order to describe its 
dynamics. These kind of resonances due to the action of the ac 
force and damping were successfully explained by using 
a more general {\it Antsatz}, the so called Generalized 
Traveling Wave Ansatz ({\it GTWA}) \cite{prl,rc,gtwa} together with the 
Rice approximated solution \cite{sal-scott,Rice}. This latter method is  
easier to apply than the former perturbation theory, however it 
%does not consider explicitly the effect of the phonons and 
is only supported in the projection technique. 
Recently, the equivalence of the {\it GTWA} and the variation 
of the momentum and the energy (two constants of motion of  
the unperturbed problem) of the nonlinear Klein-Gordon 
systems when two CCs are considered \cite{pre}, has been shown. When it is necessary 
to use more that two CCs, this equivalence 
is unclear due to the absence of (or lack of knowledge of) the 
constants of motion for the general unperturbed nonlinear 
Klein-Gordon equation (except for the integrable sG equation).  

In order to solve the perturbed nonlinear Klein-Gordon equation,  
we need to assume how  explicit the approximated solution of the problem is, i.e. the 
{\it Ansatz}.  
Usually the {\it Ansatz} involves a number of unknown variables 
(collective coordinates) and by using either 
the perturbation theory or CCs approach we obtain the equations of 
motion (system of ordinary differential equations) that the CCs satisfy. 

 The equations of the CCs have been also derived by using the  
{\it Lagrangian formalism}. In spite of its simplicity, this method 
has only been used to study the kink-antikink collisions 
in the unperturbed problem \cite{ka} and the kink-impurity 
interactions in the sG and $\phi^{4}$ models 
\cite{kivshar,kivshar2}. 
Furthermore, it has been used to analyze  
kink-antikink collisions in the damped $\phi^{4}$ and sG models driven 
by an external force, however the projection technique was used 
to 
treat the dissipation separately \cite{ka1}. 
Probably, the main difficulty in using this method was the fact that 
there is not any systematic way to construct the Lagrangian 
density when a generic perturbation is added into the 
sG or $\phi^{4}$ equations. 

 The aim of this work is to 
extend the Lagrangian formalism to the generic 
perturbed nonlinear Klein-Gordon equation in order to obtain 
the evolution equations that 
obey the CCs. We explicitly consider a damping term as a 
perturbation. In this case, 
we are able to write the Lagrangian density by introducing a new 
time variable and calculate the Lagrangian as a function of the CCs  
for a given {\it Ansatz} (see the section \ref{lagra}). Some examples 
are also considered in section \ref{lagra}, where we 
analyze perturbed sG and $\phi^{4}$ equations and we  
show the 
equivalence between the {\it GTWA} and the Lagrangian formalism 
for the nonlinear Klein-Gordon equations. 
Finally, in section \ref{conclu} 
we discuss and summarize the main results of this work.          
 
\section{Lagrangian formalism} 
\label{lagra}    

 In this section we study the approximated solution of the 
following perturbed nonlinear Klein-Gordon 
equation
\begin{eqnarray}\label{eq1}
\phi_{tt}-\phi_{xx} & = & - \frac{dU}{d\phi} - \beta \phi_{t} + 
z(x,t,\phi),
\end{eqnarray} 
where the subindex $t$ and $x$ indicate the partial derivatives with 
respect to time and space, respectively; $U(\phi)$ is the nonlinear 
Klein-Gordon potential, $\beta$ is the damping coefficient and 
$z(x,t,\phi)$ represents a generic perturbation on the system. Notice that, 
by making the change of variable in time 
$\tau=\exp(-\beta t)$ \cite{nota},   
Eq.\ (\ref{eq1}) becomes in dissipationless equation, 
\begin{eqnarray}\label{eq2}
\phi_{\tau \tau}-\frac{\phi_{xx}}{\beta^{2} \tau^{2}} & = & 
- \frac{1}{\beta^{2} \tau^{2}} \frac{dU}{d\phi} + 
\frac{\tilde{z}(x,\tau,\phi)}{\beta^{2} \tau^{2}}.
\end{eqnarray} 

%%%*******
 Let us to remark that the idea to suppress the damping in the 
nonlinear systems have been already 
used in \cite{ll}, where the authors showed that the effect of damping 
in the Landau-Lifshitz equation is only a rescaling of time by a 
complex constant.  
%%%%%*****

 Now by using the Euler-Lagrange equation \cite{lagra} 
it is not difficult 
to show that Eq.\ (\ref{eq2}) can be obtained from the 
Lagrangian 
\begin{eqnarray}\label{eq3}
L & = & \int_{-\infty}^{+\infty} \, dx {\mathcal L}=  
\int_{-\infty}^{+\infty} \, dx \left\{ \frac{1}{2} 
\phi_{\tau}^{2} - \frac{1}{2} \frac{\phi_{x}^{2}}
{\beta^{2} \tau^{2}} - \frac{U(\phi)}{\beta^{2} \tau^{2}} + 
\frac{W(x,\tau) \phi_{x}}{\beta^{2} \tau^{2}} \right\},
\end{eqnarray} 
where $\mathcal{L}$ is the Lagrangian density, and 
\begin{eqnarray}\label{eq4}
W(x,\tau) & = & - \int_{x_{0}}^{x} \tilde{z}(y,\tau,\phi) dy, 
\end{eqnarray}
%%%%%%%%%**** 
where $\phi$ in the last expression represents 
the approximated solution of Eq.\ (\ref{eq1}), i.e. the 
{\it Ansatz}. Notice that, in general the expression (\ref{eq3}) 
is an approximated 
Lagrangian density corresponding to the Eq.\ (\ref{eq2}). 
In particular, if $z(x,t,\phi)$ is not a function of the field 
$\phi$,  the Eq.\ 
(\ref{eq3}) represents an exact Lagrangian density corresponding to Eq.\ 
(\ref{eq1}). %%%%%%%%%%%******
To proceed we need to assume an approximated 
solution for Eq.\ (\ref{eq2}). The {\it Ansatz} we use involves $N$ CCs  
($\vec{Y}=Y_{1}, Y_{2}, ..., Y_{N}$) and so, by inserting it in 
the Eq.\ (\ref{eq3}) we obtain the Lagrangian as a function of 
the $N$ CCs and their derivatives with respect to the new time variable $\tau$, 
$L(\vec{Y},\vec{Y}')$. From this moment on with the prime and the dot 
we denote the time derivative in relation to $\tau$ and 
$t$, respectively.  
The next step is to derive the equations of motion for a given CC $Y_{i}$ 
($i=1, 2, ..., N$) 
by using the Lagrange equation 
\begin{eqnarray}\label{eq5}
\frac{d}{d \tau} \left(\frac{\partial L}{\partial Y'_{i}}\right) 
- \frac{\partial L}{\partial Y_{i}} & = & 0.
\end{eqnarray} 

 In particular, we will analyze two cases,  
$U(\phi)=1-\cos(\phi)$ corresponding to sG equation, and 
$U(\phi)=(1/4)(1-\phi^{2})^{2}$ to $\phi^{4}$ one. 
In both examples, we use the Rice's {\it Ansatz} 
\cite{sal-scott,Rice}.  
This approximated solution reads 
\begin{eqnarray}\label{eq6}
\phi(x,\tau) &=& 4 \, {\rm atan} \left( \exp 
\left[\frac{x-X(\tau)}{l(\tau)} \right]
\right), 
\end{eqnarray} 
for the sG and 
\begin{eqnarray}\label{eq7}
\phi(x,\tau) &=& \tanh \left[\frac{x-X(\tau)}{l(\tau)}\right], 
\end{eqnarray} 
for $\phi^{4}$ equation, where $X(\tau)$ and $l(\tau)$ represent the center 
and the width of the kink, respectively.  
Substituting Eqs.\ (\ref{eq6}) and (\ref{eq7}) in (\ref{eq3}) 
and integrating we obtain 
\begin{eqnarray}\label{eq8}
L(X,X',l,l') &=& \frac{M_{0} l_{0}}{2 l} (X')^{2} + 
\frac{\alpha M_{0} l_{0}}{2 l} (l')^{2} - \frac{M_{0}}{2 \beta^{2}
\tau^{2}} \left(\frac{l_{0}}{l} + \frac{l}{l_{0}} \right) + \\  
& \, & \frac{1}{\beta^{2} \tau^{2}} \int_{-\infty}^{+\infty} 
W(X + \theta l,\tau) \phi_{\theta} d\theta,   \nonumber
\end{eqnarray} 
where $M_{0}$ is the mass of the kink, $l_{0}$ 
represents the width of the static unperturbed 
kink and $\alpha$ is a coefficient. In particular, 
$M_{0}=8$, $l_{0}=1$ and $\alpha=\pi^{2}/12$ for the sG and 
$M_{0}=4/(3 \sqrt{2})$, $l_{0}=\sqrt{2}$ and $\alpha=(\pi^{2}-6)/12$ 
for $\phi^{4}$.  
From the  Eq.\ (\ref{eq5}) and the 
Lagrangian (\ref{eq8}),  
the equations of motion for the 
CCs $X(\tau)$ and $l(\tau)$ are given by 
\begin{eqnarray}\label{eq90}
& \, & \frac{dP}{d\tau} = -\frac{1}{\beta^{2} \tau^{2}} 
\int_{-\infty}^{+\infty} 
\tilde{z}(X + \theta l,\tau,\phi) \phi_{\theta} d\theta, \qquad 
P(\tau) = \frac{M_{0} l_{0}}{l(\tau)} X', \\
%& \, & \frac{d}{d\tau} \left(\frac{\alpha M_{0} l_{0} l'}{l} \right)+  
& \, & 
\frac{\alpha M_{0} l_{0}}{l} \left(l''-\frac{(l')^{2}}{2 l} \right)+ 
%\frac{M_{0} l_{0}}{2 l^{2}} (X')^{2} + 
\frac{P^{2}}{2 M_{0} l_{0}} + 
%\frac{\alpha M_{0} l_{0}}{2 l^{2}} (l')^{2} + 
\frac{M_{0}}{2 \beta^{2} \tau^{2} l} 
\left(\frac{l}{l_{0}} - \frac{l_{0}}{l}\right) = \nonumber \\  
& \, & \frac{1}{\beta^{2} \tau^{2}} \int_{-\infty}^{+\infty} 
\frac{\partial W}{\partial l}(X+\theta l,\tau) \phi_{\theta} 
d \theta, \nonumber 
\end{eqnarray} 
or equivalent, in a more compact form, the equations for $X(t)$ and $l(t)$ read  
\begin{eqnarray}\label{eq9} 
\hspace{-0.7cm} & \, & \frac{dP}{dt} = -\beta P(t) - 
\int_{-\infty}^{+\infty} z(X + \theta l,\tau,\phi) 
\phi_{\theta} d\theta, \qquad 
P(t)= \frac{M_{0} l_{0} \dot{X}}{l(t)}, \\
\hspace{-0.7cm} & \, & \alpha [\dot{l}^{2} - 2 \beta l \dot{l} - 2 l \ddot{l}] = 
\frac{l^{2}}{l_{0}^{2}} \left(\frac{P^{2}}{M_{0}^{2}} + 1 \right) 
-1 + \frac{2 l^{2}}{M_{0} l_{0}} \int_{-\infty}^{+\infty} 
 z(X + \theta l,\tau,\phi) \theta \phi_{\theta} 
d\theta. \nonumber
\end{eqnarray}     
This system of equations has been also 
obtained by using the {\it GTWA} or the variation of the energy, $E(t)$,  
and the momentum, $P(t)$, together with the Rice's {\it Ansatz} \cite{PhD}. 
As we will later show the Lagrangian formalism and the {\it GTWA} are 
equivalent in a more general sense, i.e. the equivalence between both methods 
is regardless of the {\it Ansatz} that we use as approximated 
solution of Eq.\ (\ref{eq1}) and also of how many 
CCs we use. Notice that, for the ac forces  
$z(x,t,\phi)=\epsilon_{1} \sin(\delta t + \delta_{0})+ 
\epsilon_{2} \sin(m \delta t + \delta_{0})$ \cite{prl,rc,prl2},  
for the parametric periodic force   
$z(x,t,\phi)=\epsilon \sin(\delta t + \delta_{0})\phi$ \cite{ejpb},  
and for the driven $z(x,t,\phi)=-\epsilon$ \cite{pre2} 
we recover the equations related with the 
resonance phenomena and the ratchet effect studied 
in these references by using either the {\it GTWA} or the variation of $E(t)$ 
and $P(t)$.  
     
 In order to establish the relation between {\it GTWA} and Lagrangian 
formalism, presented here for the nonlinear Klein-Gordon equations, 
let us assume that 
the solution of Eq.\ (\ref{eq2}) depends on the $N$ CCs, so 
$\phi(x,\vec{Y}(\tau))$ with $\vec{Y}=Y_{1}, Y_{2}, ..., Y_{N}$. 
Then, the Lagrangian density in Eq.\ 
(\ref{eq3}) is a function of $\phi(x,\vec{Y}(\tau))$, so that 
\begin{eqnarray} \label{eq11}   
L & = & \int_{-\infty}^{+\infty} \, dx \, 
{\mathcal L}(\phi(x,\vec{Y}(\tau))). 
\end{eqnarray}
 By inserting Eq.\ (\ref{eq11}) in (\ref{eq5}), after some straightforward calculations 
(see e.g. \cite{ka1}a) we obtain 
\begin{eqnarray} \label{eq12}   
& \, & \int_{-\infty}^{+\infty} \, dx \, \left\{\frac{\partial}{\partial \tau} 
\left(\frac{\partial {\mathcal L}}{\partial \phi_{\tau}} \right) + 
\frac{\partial}{\partial x} 
\left(\frac{\partial {\mathcal L}}{\partial \phi_{x}} \right) - 
\frac{\partial {\mathcal L}}{\partial \phi}
 \right\} \frac{\partial \phi}{\partial Y_{i}}. 
\end{eqnarray}
This equation is just the projection of Eq.\ (\ref{eq2}) into the functions $\partial \phi/\partial Y_{i}$. 
%Without loss of generality we will use 
%$\phi(x-Y_{1},Y_{2},...,Y_{N})$ instead of 
%$\phi(x,Y_{1},Y_{2},...,Y_{N})$, where $Y_{1}$ represents the 
%center of mass of the kink and $Y_{j}$ (j=2, 3, ..., N) 
%are the others $N-1$ CC. 
By substituting the {\it Ansatz} $\phi(x,\vec{Y})$ 
in Eq.\ (\ref{eq12})  
we obtain the following set of $N$ 
equations for the $N$ CCs by using the Lagrangian density defined in (\ref{eq3}):
\begin{eqnarray} \nonumber   
%& \, & 
%\int_{-\infty}^{+\infty} \, dx \, 
%\left\{
%\frac{\partial \phi}{\partial Y_{1}}\frac{\partial \psi}{\partial Y_{1}^{'}}  Y_{1}^{''} + 
%\frac{\partial \phi}{\partial Y_{1}}\frac{\partial \psi}{\partial Y_{i}^{'}} Y_{i}^{''} + 
%\left[\frac{\partial \phi}{\partial Y_{1}}\frac{\partial \psi}{\partial Y_{i}} - 
% \frac{\partial \psi}{\partial Y_{1}}\frac{\partial \phi}{\partial Y_{i}}\right]
%Y_{i}^{'} + \right. \\
%\qquad & \, & \left. \frac{1}{\beta^{2} \tau^{2}} \left[
%\frac{\partial \phi}{\partial Y_{1}} \frac{\delta H}{\delta \phi} - 
%\beta \tau \frac{\partial \psi}{\partial Y_{1}} \frac{\delta H}{\delta \psi} \right]
%-\frac{\tilde{z}}{\beta^{2} \tau^{2}} \frac{\partial \phi}{\partial Y_{1}}
%\right\}
% = 0,  \label{eq13}   \\
 \nonumber 
& \, & 
\int_{-\infty}^{+\infty} \, dx \, 
\left\{
%\frac{\partial \phi}{\partial Y_{k}} 
%\frac{\partial \psi}{\partial Y_{1}^{'}}  Y_{1}^{''} + 
\frac{\partial \phi}{\partial Y_{k}} 
\sum_{i=1}^{N} \frac{\partial \tilde{\psi}}{\partial Y_{i}^{'}} Y_{i}^{''} + 
\sum_{i=1 (i\ne k)}^{N}\left[
\frac{\partial \phi}{\partial Y_{k}}
\frac{\partial \tilde{\psi}}{\partial Y_{i}}- 
\frac{\partial \tilde{\psi}}{\partial Y_{k}}\frac{\partial \phi}{\partial Y_{i}}
\right]
Y_{i}^{'} + \right. \\
\qquad & \, & \left. \frac{1}{\beta^{2} \tau^{2}} \left[
\frac{\partial \phi}{\partial Y_{k}} \frac{\delta H}{\delta \phi} - 
\beta \tau \frac{\partial \tilde{\psi}}{\partial Y_{k}} 
\frac{\delta H}{\delta \psi} \right]
-\frac{\tilde{z}}{\beta^{2} \tau^{2}} \frac{\partial \phi}{\partial Y_{k}}
\right\}
 = 0, \label{eq14}  
\end{eqnarray}
where $k=1, 2, ..., N$, $\tilde{\psi}\equiv \phi_{\tau} = 
\tilde{\psi}(x,\vec{Y},\vec{Y}^{'})$ and $H$ is the Hamiltonian 
corresponding to Eq.\ (\ref{eq1}) with $\beta=0$ and $z=0$. In the variable $t$ this system of $N$ 
equations reads 
\begin{eqnarray} 
%\nonumber   
%& \, & 
%\int_{-\infty}^{+\infty} \, dx \, 
%\left\{
%\frac{\partial \phi}{\partial Y_{1}}\frac{\partial \psi}{\partial \dot{Y}_{1}}  \ddot{Y}_{1} + 
%\frac{\partial \phi}{\partial Y_{1}}\frac{\partial \psi}{\partial \dot{Y}_{i}} \ddot{Y}_{i} + 
%\left[\frac{\partial \phi}{\partial Y_{1}}\frac{\partial \psi}{\partial Y_{i}} - 
% \frac{\partial \psi}{\partial Y_{1}}\frac{\partial \phi}{\partial Y_{i}}\right]
%\dot{Y}_{i} + \right. \\
%\qquad & \, & \left.  \frac{\partial \mathcal H}{\partial Y_{1}} - (z-\beta \psi) 
%\frac{\partial \phi}{\partial Y_{1}}
%\right\}
% = 0,  \label{eq15}   \\
 \nonumber 
& \, & 
\int_{-\infty}^{+\infty} \, dx \, 
\left\{
\frac{\partial \phi}{\partial Y_{k}}
\sum_{i=1}^{N} \frac{\partial \psi}{\partial \dot{Y}_{i}}  \ddot{Y}_{i} + 
\sum_{i=1 (i \ne k)}^{N} 
\left[\frac{\partial \phi}{\partial Y_{k}} \frac{\partial \psi}{\partial Y_{i}}
 -\frac{\partial \psi}{\partial Y_{k}} 
\frac{\partial \phi}{\partial Y_{i}}\right]
\dot{Y}_{i} + \right. \\
\qquad & \, & \left. 
 \frac{\partial \mathcal H}{\partial Y_{k}} - (z-\beta \psi) 
\frac{\partial \phi}{\partial Y_{k}}
\right\}
 = 0, \label{eq16}  
\end{eqnarray}  
where $\mathcal H$ is the Hamiltonian density of the unperturbed 
Eq.\ (\ref{eq1}) [put in (\ref{eq1}) 
$\beta$ and $z$ equal to zero]. This set of equations for the $N$ CCs can be obtained 
also by using the {\it GTWA}. 
This set of equations is reduced when two CCs, 
$Y_{1} \equiv X(t)$ and $Y_{2} \equiv l(t)$, are considered.  
In this case we recover the Eqs.\ (7) and (10) of \cite{pre}, obtained 
by using either the 
{\it GTWA} or the variation of the energy and the momentum. 
%for two CCs, $Y_{1} \equiv X(t)$ and $Y_{2} \equiv l(t)$.   
  
\section{Conclusions} \label{conclu}
         
 In this work, using the Lagrangian formalism, we have developed 
an alternative method  
in order to obtain the system of ODEs that satisfy the CCs in 
the generic perturbed nonlinear Klein-Gordon  
models. The main advantage of this method  
over previous ones is that it involves less 
calculations, so that the equations of 
motion for the CCs can be derived straightforwardly. 
Indeed, given the generic perturbed nonlinear  
Klein-Gordon equation (\ref{eq1}), we have shown that 
with a nonlinear change of the time variable  
in (\ref{eq1}) it is possible 
to write the dissipationless equation (\ref{eq2}) and its Lagrangian density 
(\ref{eq3}). %%%%%%%%*******
(In general $\mathcal{L}$ is an approximated Lagrangian density 
of (\ref{eq1}). In particular, when the 
perturbation $z(x,t,\phi)$ does not depend on $\phi$, we are able to obtain an 
exact Lagrangian density corresponding to Eq.\ (\ref{eq1})). 
%The idea to supress the damping by a time change of variables was applied 
%in \cite{ll} for the Landau-Lifshitz equation.  %%%%%%%%%%%*****
Furthermore, given an {\it Ansatz} for the solution of (\ref{eq1}) we 
can, first, calculate the Lagrangian as a function of the $N$ CCs by using 
the Lagrangian density defined by Eq.\ (\ref{eq3}). Then, from the Lagrange 
equation (\ref{eq5}) we can obtain the corresponding system of 
ODEs for the collective variables.  

%As in the previous methods in order to calculate explicitly 
%the Lagrangian as a function of CC,  
%it is necessary to assume an {\it Ansatz}. 
By using the Rice 
{\it Ansatz} \cite{sal-scott,Rice} 
we have obtained the system of ODEs that satisfy the collective variables for the 
perturbed sine-Gordon and $\phi^4$ systems. 
These equations coincide with those obtained by the  
{\it GTWA} \cite{gtwa} or the variation of energy and the 
momentum \cite{pre}. 
In particular, for ac and dc forces we have recovered the results 
obtained in \cite{prl,rc,prl2,ejpb,pre2}. 
We would like to remark that this approach, presented here 
for the perturbed sG and $\phi^{4}$ models with the Rice {\it Ansatz} can be 
extended %without any difficulty 
to others perturbed nonlinear Klein-Gordon 
systems, for example, DsG and 
ADSG equations \cite{sn}. %%%%%%%%%%%%%%%%%************
Furthermore, the Lagrangian formalism 
is not restricted to one dimensional system. Indeed, it recently 
have been applied  to solve approximately a problem concerning 
to the vortex theory \cite{zago}. 
%%%%%%%%%%%%%%%%%**************  
  
Finally for $N$ CCs, 
we have shown the equivalence between this method and the 
{\it GTWA}, regardless of the {\it Ansatz} that we use  
for the approximated solution of Eq.\ (\ref{eq1}).

%This method allow us to consider the effect of radiation field 
%(by including more CC) in 
%a more easy way that the perturbation theory since the projection 
%tecnique is not required.  

\section{Acknowledgments} 
We would like to thank Renato \'Alvarez-Nodarse and Franz Mertens for the useful 
discussion on this work.  
This work has been
supported by the Ministerio de Ciencia y Tecnolog\'\i a of Spain 
through grants BFM2001-3878-C02 and 
by the Junta de Andaluc\'{\i}a under the
project FQM-0207.

\end{document}